\def \date         {\ifcase\month \message{zero} \or
                    January \or February \or March \or April \or May \or June 
                    \or July \or 
                    August \or September \or October \or November \or 
                    December \fi
                    \space\number\day, \number\year}
\def \eg           {{e.g.}}
\def \etal         {{et~al. }}
\def \h2         {\hbox{H$_2$}}

\def \eng          {\hbox{ergs$\,$s$^{-1}$}}
\def\approxlt{\lower.2em\hbox{$\buildrel < \over \sim$}}
\def\approxgt{\lower.2em\hbox{$\buildrel > \over \sim$}}

\def \ha           {H$\alpha$} 
\def \kms          {\hbox{km$\,$s$^{-1}$}}

\documentclass{emulateapj}
\begin{document}

\title{Non-nuclear Hyper/Ultraluminous X-ray Sources in the Starbursting Cartwheel Ring Galaxy} 

\author{Yu Gao\altaffilmark{1}, Q. Daniel Wang\altaffilmark{1}, 
P. N. Appleton\altaffilmark{2}, and Ray A. Lucas\altaffilmark{3}}

\altaffiltext{1}{University of Massachusetts, Department of Astronomy,
LGRT-B 619E, 710 North Pleasant Street, Amherst, MA 01003-9305}
\altaffiltext{2}{SIRTF Science Center, Caltech, MS 220-6, 
1200 E. California Blvd.,Pasadena, CA 91125}
\altaffiltext{3}{Space Telescope Science Institute, 3700 San Martin Drive,
Baltimore, MD 21218}

\accepted{2003 September 5}

\begin{abstract}

We report the Chandra/ACIS-S detection of more than 20 ultraluminous 
X-ray sources (ULXs, $L_{\rm 0.5-10 keV} \approxgt 3\times 10^{39} \eng$)
in the Cartwheel collisional ring galaxy system, of which over 
a dozen are located in the outer active star-forming ring. 
A remarkable hyperluminous X-ray source (HLX, 
$L_{\rm 0.5-10 keV} \approxgt 10^{41} \eng$ assuming isotropic radiation),
which dominates the X-ray emission from the Cartwheel ring, is located 
in the same segment of the ring as most ULXs. These powerful H/ULXs appear 
to be coincident with giant HII region complexes, young star clusters, 
and radio and mid-infrared hot-spots: all strong
indicators of recent massive star formation.  The X-ray spectra show
that H/ULXs have similar properties as those of the {\it most luminous}
ULXs found in the nearest starbursts and galaxy mergers such as the
Antennae galaxies and M82.  The close association between the X-ray
sources and the starbursting ring strongly suggests that the H/ULXs
are intimately associated with the production and rapid evolution of
short-lived massive stars.  The observations represent the most
extreme X-ray luminosities discovered to date associated with star-forming
regions---rivaling the X-ray luminosities usually associated with
active galactic nuclei.

\end{abstract}


\section{Introduction}

The high resolution capabilities of the Chandra have led to the
discovery of a large population of extraordinarily X-ray luminous
point-like non-nucleus sources in many nearby galaxies (\eg, Fabianno,
Zezas, \& Murray 2001; Colbert et al. 2003). These so called
ultraluminous X-ray sources (ULXs; Makishima et al. 2000) can have
apparent isotropic broad-band X-ray luminosities of $L_{\rm 0.5-10 keV}$:
hundreds of times the Eddington limit for a neutron star or a stellar
mass black hole.  Although rarely present in normal galaxies, very 
luminous ULXs with $L_{\rm 0.5-10 keV}
\approxgt 3\times 10^{39} \eng$ are often found in starbursts and
IR-luminous galaxy mergers.  For example, in both the Antennae
galaxies and the luminous infrared (IR) galaxy merger NGC~3256, more than
half a dozen such luminous ULXs are detected (Fabianno et al. 2001; Lira \etal
2002), and numerous less luminous point-like sources are spread
over the merging disks (Zezas et al. 2002). The Chandra observations
of the Cartwheel galaxy\footnote{ $H_0$=75\kms~Mpc$^{-1}$ 
($cz$=9090\kms, $d_{\rm L}$=122~Mpc) is used in this paper} 
presented here show an unusually large number ($>$20) of high-luminosity
ULXs, most of which appear highly correlated with the narrow,
young, outer starburst ring.

Hyperluminous X-ray sources (HLXs, $L_{\rm 0.5-10 keV} \approxgt
10^{41} \eng$, Matsumoto et al. 2001; Kaaret \etal 2001) are 
intriguing as they are apparently more luminous than the entire X-ray 
luminosity of a normal galaxy---with luminosities approaching 
that of luminous active galactic nuclei (AGNs). H/ULXs appear variable
(Fabianno et al. 2003)---their peak luminosity often changing by 
more than an order of magnitude within months to years 
(Matsumoto et al. 2001; Strickland \etal 2001).  Here, we
report an extraordinary HLX, as well as more than 
a dozen ULXs, located within and along the same portion of 
the active star-forming outer ring of the Cartwheel.

The Cartwheel galaxy has been studied in the radio (Higdon
1996), IR (Marcum, Appleton, \& Higdon 1992; Charmandaris et al. 1999),
optical (Hidgon 1995; Struck et al. 1996; Amram et al. 1998), and X-rays
(Wolter, Trinchieri, \& Iovino 1999).  The crisp outer ring is
believed to have been created as stars formed in radially-expanding
density waves caused by a companion (``intruder'') galaxy plunging 
through the center of a gas-rich disk (see Lynds \& Toomre 1976;
Toomre 1977; and review by Appleton \& Struck 1996). The outer
ring appears to have propagated into a low-metallicity region 
of the disk (Fosbury \& Hawarden 1977) where it has triggered 
recent active star formation. 

Both \ha~ and radio continuum observations (Higdon 1995, 1996)
indicate that the dominant star-forming sites are in the outer ring of
the southern quadrant, which constitutes about 80\% of the total
emission from the entire galaxy.  Chandra observations discussed here
clearly show that the dominant X-ray emission originates 
in the same starbursting southern ring quadrant.

\section{Imaging Analysis \& Spectral Fitting}

The Chandra data analysis of the Cartwheel is part
of a systematical, uniform, archival study of 
nearly 20 interacting/merging galaxies.
The Cartwheel was observed by ACIS-S (OBSID: 2019, PI: A. Wolter) 
with a 75 ksec exposure from 2001 May 26 to 27. Data calibration
was performed with CIAO v2.3 using the latest calibration
database. Events files of level one were obtained from the Chandra
archive, and were corrected for the aspect offset before processing
into level two. Bad pixels and background flares as well as
streaks were removed, and final clean maps were created. 
Further data analysis and image processing were done in IDL. This 
includes source detection, images of individual bands, and image 
smoothing (see Wang, Chaves, \& Irwin 2003). We also followed 
various CIAO threads to extract 
the X-ray spectra for point-like and extended sources. Spectral
fitting was done with XSPEC. We also ran the timing analysis, but none of
the point-like sources showed detectable variability 
within the observing time frame.  


Fig.~1a shows broad-band X-ray contours overlaid on an HST/WFPC2
B-band image. Almost all the X-ray
emission in the Cartwheel originates from point-like sources within the
southern quadrant of the outer ring. The sources are nearly coincident 
with the strong \ha~, radio continuum emission and blue super-star 
clusters (SSCs). Other portions of the outer ring show little X-ray 
emission except where exceptionally bright SSCs are seen. We label all
the point-like sources in the immediate surroundings of the Cartwheel
in Fig.~1b and detail them in Table~1. 


The companion galaxy G1 (spiral) contains 6
point-like X-ray sources, and the early-type spiral G2 is seen as a
fainter diffuse source (Fig~1). The farthest companion galaxy G3
is also significantly detected, with one ULX in the eastern
edge of its disk. In addition, a faint, diffuse X-ray envelope which
includes the Cartwheel, G1 and G2 is marginally detected.

The absence of any point-like X-ray source in the nuclear region of the
Cartwheel rules out the existence of AGN.
Although rather complex optical structures (Struck \etal 1996), 
and prominent mid-IR emission (Charmandaris \etal 1999) in 
the inner disk/ring, the X-ray emission from Cartwheel's inner disk, 
including the nucleus, is extremely weak.


Table~2 summarizes our spectral fitting. For the strongest, source 11, 
both an absorbed Raymond-Smith (RS) and Mekal thermal plasma models 
failed to yield the acceptable fit, giving an unrealistically high 
temperature, $\approxgt 10$ keV, for the plasma. The
absorbed power-law (PO) and multicolor accretion disk (MCD) models are
both acceptable, but the data show excess emission features around 
1.4~keV. This appears to be similar to the line features (Mg~$XI$ and
Mg~$XIII$) observed in a few ULXs in the Antennae (Zezas
\etal 2002). We thus modeled the spectrum with an additional narrow
Gaussian (Gau) component to the PO (Fig.~2a) and MCD models. Both can 
give improved fit to the data. 


Assuming isotropic emission, the absorption-corrected hard and
broad-band X-ray luminosities $L_{\rm 2-10 keV}$ and $L_{\rm 0.5-10 keV}$
of source 11 are $\sim$ 0.6--0.9 and 0.9--1.3~$\times
10^{41}\eng$ respectively. Thus, this source is an HLX, and 
could have a total luminosity 
$L_{\rm 0.05-100 keV}$ as large as $\sim 5.0\times 10^{41}\eng$.
In comparison, ROSAT data suggest an intrinsic 
$L_{\rm 0.5-5 keV} \sim 2.3 \times 10^{41}\eng$ for the detected 
outer ring assuming $N_{\rm H}=2\times 10^{21}$cm$^{-2}$ 
(Wolter et al. 1999), and most of the emission detected by ROSAT
is presumably from source 11.

The whole southern ring, including diffuse emission and all the 
sources, taken together, can be fit well
with an absorbed PO+RS model. So can the diffuse emission 
of the entire system (Fig.~2b,c, Table~2). 
The un-absorbed hard and broad-band X-ray luminosities of the south ring 
are 1.1 and $3.8\times 10^{41}\eng$. And the total luminosity
$L_{\rm 0.05-100 keV}$ is as large as $1.2 \times 10^{42}\eng$.
Therefore, on
average, all detected point-like sources in the south ring would
truly be ULXs, more luminous than the {\it most luminous} ULXs 
($L_{\rm 0.5-10 keV} \sim 6\times 10^{39}\eng$) detected in the 
Antennae. Although detailed spectral analysis is difficult for 
each of these point-like sources owing to the limited count 
statistics ($\sim 100$ photons), we can roughly estimate
the X-ray flux according to their count rates.  For instance, each of
the double-ULX (sources 2 \& 3) in the northwest end of the south 
ring has a count rate only 3 times smaller than that of the HLX. Both of 
the double-ULX (sources 15 \& 17) southeast of the HLX also 
have count rates only a factor of 5 lower than that of the HLX.  Thus, 
in combination, these 4 ULXs might have X-ray luminosity comparable
to that of the HLX. The faintest sources detected have a luminosity of
$L_{\rm 0.5-10 keV} \sim 3\times 10^{39}\eng$.

A point-like source 31, $\sim 10$~kpc north of G2, is 
likely a background galaxy or AGN as it has a faint optical 
counterpart in the HST image. Limited spectrum can be fit by 
an absorbed PO with a photon index $\sim 1.9$, but a much less 
absorption column density $N_{\rm H} \sim 0.5\times 10^{21}$cm$^{-2}$.

\section{Active Star-forming Ring Knots and the H/ULXs}

Most H/ULXs are located within the portion
of the Cartwheel ring which is experiencing the most powerful 
current star formation. 
Few ULXs discovered so far have prominent optical counterparts
(\eg, Immler \etal 2003; Wu \etal 2001), but almost all H/ULXs 
in the Cartwheel are closely associated with giant complexes of 
HII regions and blue SSCs (Fig.~1a). 

Although the HLX lies within $\sim 10''$ of the strongest \ha~
knot, CW-17, which is also the hot-spot in radio and
mid-IR (Higdon 1995; Charmandaris \etal 1999), it
is not coincident with this strongest starburst. Rather the HLX 
lies closer ($ 1.7''\sim 1$~kpc)
to the fainter \ha~ feature CW-20. In contrast, CW-17 coincides 
with a much softer (and fainter) ULX source 12.
Hence the HLX defies an absolute one-to-one correspondence with an
optical knot, although it lies within the arc of bright 
\ha~ and radio emission which
characterizes this portion of the ring.

We also give cross-identifications from the position match
between H/ULXs and \ha~ knots in Table~1.
All 5 strongest \ha~ knots (CW-14, 15, 17, 24, 25) with
\ha~ luminosity $\approxgt 1.3\times 10^{41}\eng$), 
20 times more luminous than 30 Doradus (Wang 1999), 
are coincident with the H/ULXs within 1$''.7$. Moreover,
except three ULXs (sources 10, 20, \& 21) interior to, and others 
(e.g., source 16 and those near G1, G2) outside of 
the outer ring, all H/ULXs seem to have spatial correspondence 
with the ring of giant HII region complex.
Nevertheless, a number of fairly strong HII regions do not 
have corresponding X-ray sources. 

The evidence that, 1) HLX lies close to an HII region, 2) all five of the
strongest \ha~ knots have corresponding ULXs, 3) the large
number of matches between \ha~ knots and other ULXs in the narrowly 
defined ring, leads to the conclusion that H/ULXs 
appear to be directly linked to the production and evolution of 
the short-lived massive stars in the Cartwheel.

\section{Nature of H/ULXs}

The Cartwheel is the record holder in hosting both the HLX and the
largest number of the {\it most luminous} ULXs in one galaxy. It has
been argued on observational and theoretical grounds (Appleton \&
Struck-Marcell 1996; Bransford et al. 1998) that the
triggering of newly-formed stars in ring galaxies occurs approximately
simultaneously as the wave propagates out through the disk---the outer
ring representing the most recently formed stars, with representative
ages $<$ 10$^{7}$ yrs. In this picture, the ring represents the outermost
progress of a wave that began at the disk-center some 300 Myrs
previously, created by the central perturbation of the intruder, 
either G3 or G1 (Mihos \& Hernquist 1994; Higdon 1996; 
Struck et al. 1996).  The
striking similarity between the X-ray source distribution and the
young SSCs and HII regions suggests a strong causal connection between
them. The lack of radial spread in the ring X-ray sources (with the
exception of the three interior sources) indicates that, like the
star-forming ring, the X-ray sources are linked to the active
star formation episode and their young ($<$ 10 Myrs) ages.

The two most likely sources of X-ray emission associated with massive
young star-forming regions are probably supernovae (SNe) or extremely 
young SN remnants (SNRs) and the high-mass X-ray binaries
(HMXBs). We can almost rule-out low-mass X-ray binaries (LMXBs)
to be the significant sources for H/ULXs along the Cartwheel
narrow ring, although intermediate-mass black holes (IMBHs, 
see review by Miller \& Colbert 2003)
are likely viable. It is conceivable that LMXBs and/or background
sources could be responsible for the three ULXs interior to the
ring. Three ``ULXs'' outside the Cartwheel with faint optical
counterparts are likely background galaxies. We restrict ourselves
here to the majority of H/ULXs in the ring.

The interaction between the expanding SN ejecta and the dense 
circumstellar medium of the progenitor massive star can produce
high X-ray luminosity $\sim 10^{40}\eng$ (\eg, Pooley \etal 2002). 
Therefore, such young ``SNRs'' are not like classical Cas-A and are
extremely bright with hard X-ray spectra. 
Some ULXs seem close to being resolved by Chandra at
$0.5''$ (300 pc) resolution, suggesting that each ULX may
be a composite of several individual luminous SNRs. If we adopt a
SN rate for the outer ring of 0.1-1 yr$^{-1}$ derived from the
non-thermal radio continuum (Higdon 1996), and a minimum age for
the ring HII regions of 1~Myrs, then we would predict a few 
times 10$^5$ SNRs created over this period. Assuming that the 
X-ray phase of young SNRs is very short-lived ($\approxlt 10^2$ 
yrs), then we might predict a few tens of young SNRs in the 
outer ring based on the radio flux alone. The individual ULXs in 
the ring could plausibly be collections of several such young SNRs. 
ULXs 6--8, 12--15 \& 17 (Fig.~1b) may be in this category,
being closely associated with radio hot-spots. Source 21,
which lies just inside the outer ring, has no H$\alpha$ emission, 
but has radio emission, may also be associated with SNRs.

In contrast, the HLX has only weak extended emission in both \ha~ 
and radio continuum. Its extreme X-ray luminosity does not seem 
to favor young SNRs unless they are unusually bright. 
Spectral fitting (Fig.~2a) also does not
support a SNR scenario for the HLX, given both absorbed 
PO and MCD models could fit the data, whereas the thermal plasma 
models failed. This source may be the best candidate for a HMXB 
and/or IMBH since the formation of massive black hole in dense star
clusters is possible (\eg, Rasio, Freitag, \& G\"urkan 2003).

\newpage

\begin{deluxetable}{lrrrr}
  \tabletypesize{\footnotesize}
  \tablecaption{Point-like X-ray Sources (H/ULXs) in the Cartwheel Galaxy System}
  \tablewidth{0pt}
  \tablehead{
  \colhead{No.} &
  \colhead{CXO Name} &
  \colhead{CR $({\rm~cts~ks}^{-1})$} &
  \colhead{\ha~ ID\tablenotemark{a}} &
  \colhead{Offset ($''$)}
  }
  \startdata

   1 &  J003737.56-334342.7 & $     0.13  \pm   0.05$ & --    & --  \\
   2 &  J003737.59-334257.3 & $     2.11  \pm   0.21$ & CW-26 & 2.5 \\
   3 &  J003737.61-334255.8 & $     1.91  \pm   0.19$ & CW-27 & 3.1 \\
   4 &  J003737.87-334253.6 & $     0.16  \pm   0.06$ & --    & --  \\
   5 &  J003738.36-334309.3 & $     0.24  \pm   0.07$ & CW-25 & 3.0 \\
   6 &  J003738.75-334316.7 & $     1.28  \pm   0.15$ & CW-24 & 1.4 \\
   7 &  J003738.82-334319.1 & $     0.72  \pm   0.12$ & CW-23 & 0.7 \\
   8 &  J003738.97-334317.5 & $     0.16  \pm   0.06$ & --    & --  \\
   9 &  J003739.16-334230.9 & $     0.60  \pm   0.10$ & CW-29 & 1.3 \\
  10 &  J003739.22-334250.6 & $     0.96  \pm   0.13$ & --    & --  \\
  11 &  J003739.40-334323.7 & $     5.75  \pm   0.31$ & CW-20 & 1.7 \\
  12 &  J003740.26-334327.5 & $     0.28  \pm   0.07$ & CW-17 & 1.4 \\
  13 &  J003740.46-334325.4 & $     0.24  \pm   0.07$ & --    & --  \\
  14 &  J003740.74-334330.7 & $     0.35  \pm   0.08$ & CW-15 & 0.5 \\
  15 &  J003740.88-334331.3 & $     1.04  \pm   0.13$ & CW-14 & 1.7 \\
  16 &  J003741.06-334221.9 & $     0.31  \pm   0.08$ & --    & --  \\
  17 &  J003741.09-334332.3 & $     1.26  \pm   0.16$ & CW-12 & 1.6 \\
  18 &  J003741.73-334235.7 & $     0.15  \pm   0.06$ & CW-3/4  & 2.3/2.7 \\
  19 &  J003742.01-334326.8 & $     0.20  \pm   0.07$ & CW-10/9 & 1.8/0.7 \\
  20 &  J003742.15-334314.2 & $     0.31  \pm   0.09$ & --    & --  \\
  21 &  J003742.50-334304.5 & $     0.33  \pm   0.08$ & --    & --  \\
  22 &  J003742.80-334212.8 & $     0.67  \pm   0.12$ & --    & --  \\
  23 &  J003742.87-334210.2 & $     0.32  \pm   0.08$ & --    & --  \\
  24 &  J003742.96-334204.6 & $     0.69  \pm   0.12$ & --    & --  \\
  25 &  J003743.02-334206.3 & $     0.84  \pm   0.13$ & --    & --  \\
  26 &  J003743.13-334143.3 & $     0.23  \pm   0.07$ & --    & --  \\
  27 &  J003743.14-334204.4 & $     1.13  \pm   0.15$ & --    & --  \\
  28 &  J003743.70-334147.6 & $     0.25  \pm   0.07$ & --    & --  \\
  29 &  J003743.87-334210.2 & $     0.67  \pm   0.11$ & --    & --  \\
  30 &  J003745.32-334229.0 & $     1.13  \pm   0.15$ & --    & --  \\
  31 &  J003745.63-334152.3 & $     4.91  \pm   0.33$ & --    & --  \\

\tablenotetext{a}{\ha~ knots as identified by Higdon (1995) with a
point-spread function of $1.''7$ FWHM.}

\enddata

 \end{deluxetable}

\begin{deluxetable}{lrrrlll}
\scriptsize
\tablecolumns{7}
\tablecaption{Spectral Fitting and Measurements of the Cartwheel}
\tablehead{
\colhead{Source}       &     \colhead{$N_{\rm H}$}     &
\colhead{Model Parameters\tablenotemark{a}}  &  \colhead{$\chi^2/d.o.f.$}  &
\colhead{$L_{\rm 2-10 keV}$}   &   \colhead{$L_{\rm 0.5-10 keV}$}   &
\colhead{$L_{\rm 0.05-100 keV}$}   \\
\colhead{   }        &     \colhead{10$^{21}$cm$^{-2}$}    &
\colhead{$\Gamma$, T\tablenotemark{b}} & \colhead{ }    & 
\colhead{ }    &   \colhead{10$^{41}$\eng}    &  
\colhead{ } }

\startdata

  HLX   &  3.5$\pm$.7&1.6$\pm$.2, .. (PO) & 17.5/21 & 0.8 & 1.3 & 4.1 \nl
  HLX   &  2.0$\pm$.5&.., 1.5$\pm$.2 (MCD) & 16.4/21 & 0.6 & 0.9 & 1.0 \nl
  HLX   &  2.6$\pm$.9&1.4$\pm$.2, 1.4$\pm$.1 (PO+Gau)& 11.2/18 & 0.9 & 1.2 & 5.5 \nl
  HLX   &  1.6$\pm$.5&.., 1.7$\pm$.4/1.4$\pm$.1 (MCD+Gau) & 9.5/18 & 0.6 & 0.9 & 1.0 \nl
  South Ring& 3.7$\pm$1.9 &2.2$\pm$.3, 0.18$\pm$.02 (PO+RS) & 22.0/23 & 1.1 & 3.8 & 12.5 \nl
  All\tablenotemark{c} & 1.8$\pm$2.6 & 2.3$\pm$.9, 0.23$\pm$0.15 (PO+RS) & 6.2/12 & .24 & .53 & 2.6 \nl

\tablenotetext{a}{PO, RS, Gau, and MCD, are the power-law, 
Raymond-Smith thermal plasma, Gaussian emission line, and 
Multicolor accretion disk models, respectively.}
\tablenotetext{b}{$\Gamma$ is the photon index, various temperature 
 T is in units of keV.}
\tablenotetext{c}{The diffuse emission from Cartwheel, G1 and G2 and 
between them. But all point-like sources are excluded.}

\enddata
\end{deluxetable}

\newpage

\figcaption{(a) Broad-band X-ray contours overlaid on the HST/WFPC2 
optical image. The lowest contours are 0.0345, 0.0431~cts/pixel 
(pixel$\sim 0.5''$),
and then increase successively
by a factor of 2. (b) Soft (0.3--1.5 keV) X-ray image overlaid 
with the hard X-ray contours (1.5--7 keV) with sources labeled.
Companion galaxies G1 and G2 are labeled too. The third companion
galaxy G3 is $\sim 3'$ in northeast, outside the region shown in
the figure.\label{fig-1}}
 

\newpage

\figcaption{The extracted X-ray spectra of: 
(a) The HLX and fit with the absorbed power-law (PO) plus Gaussian 
line features at 1.4 keV (top). (b) The entire ring of southern quadrant 
(middle). (c) The diffuse emission of the entire system with point-like 
sources subtracted (bottom). The fit in (b) and (c) are
the absorbed PO plus Raymond-Smith thermal plasma 
models (Table~2). \label{fig-2}}

\plotone{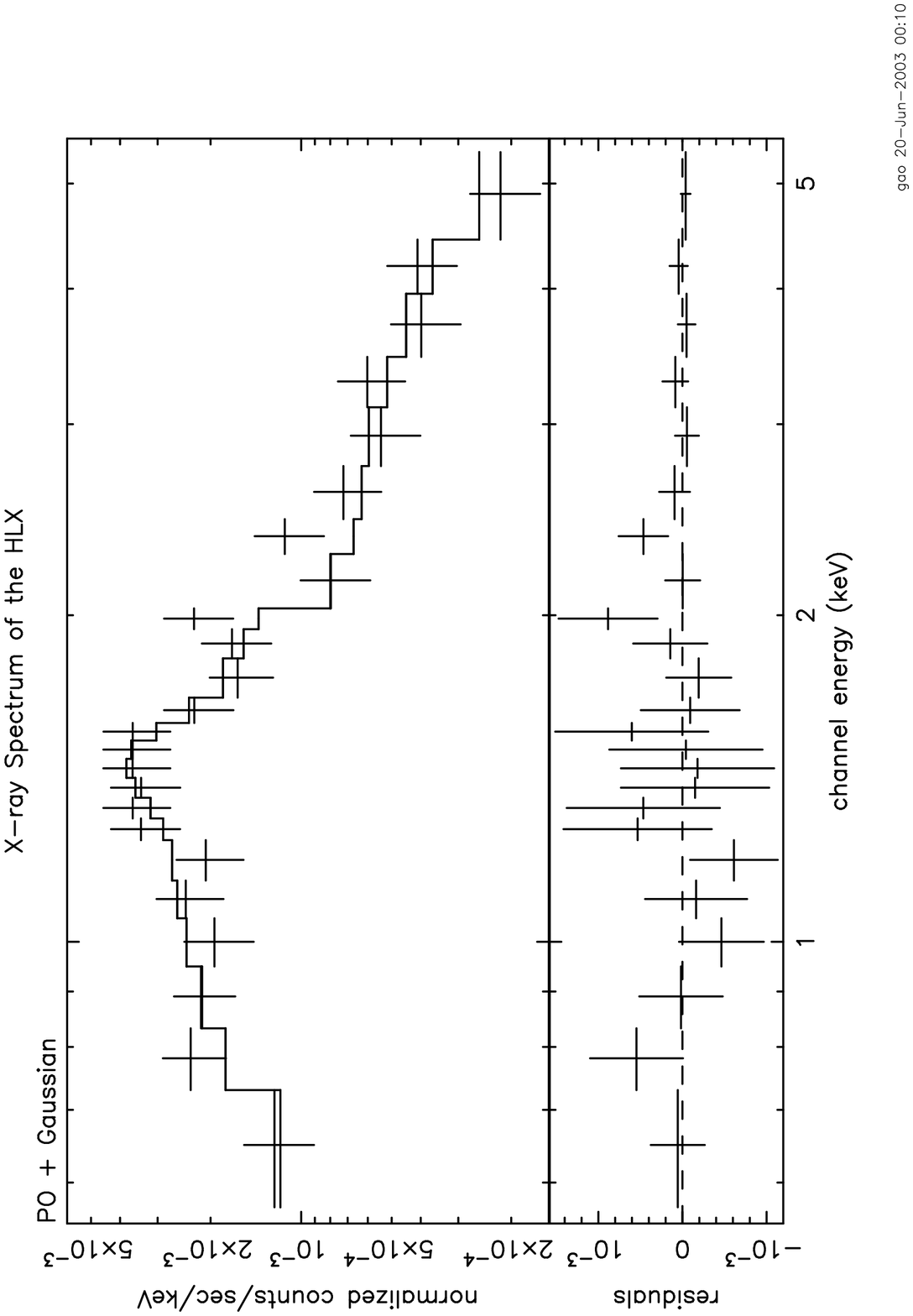}
\plotone{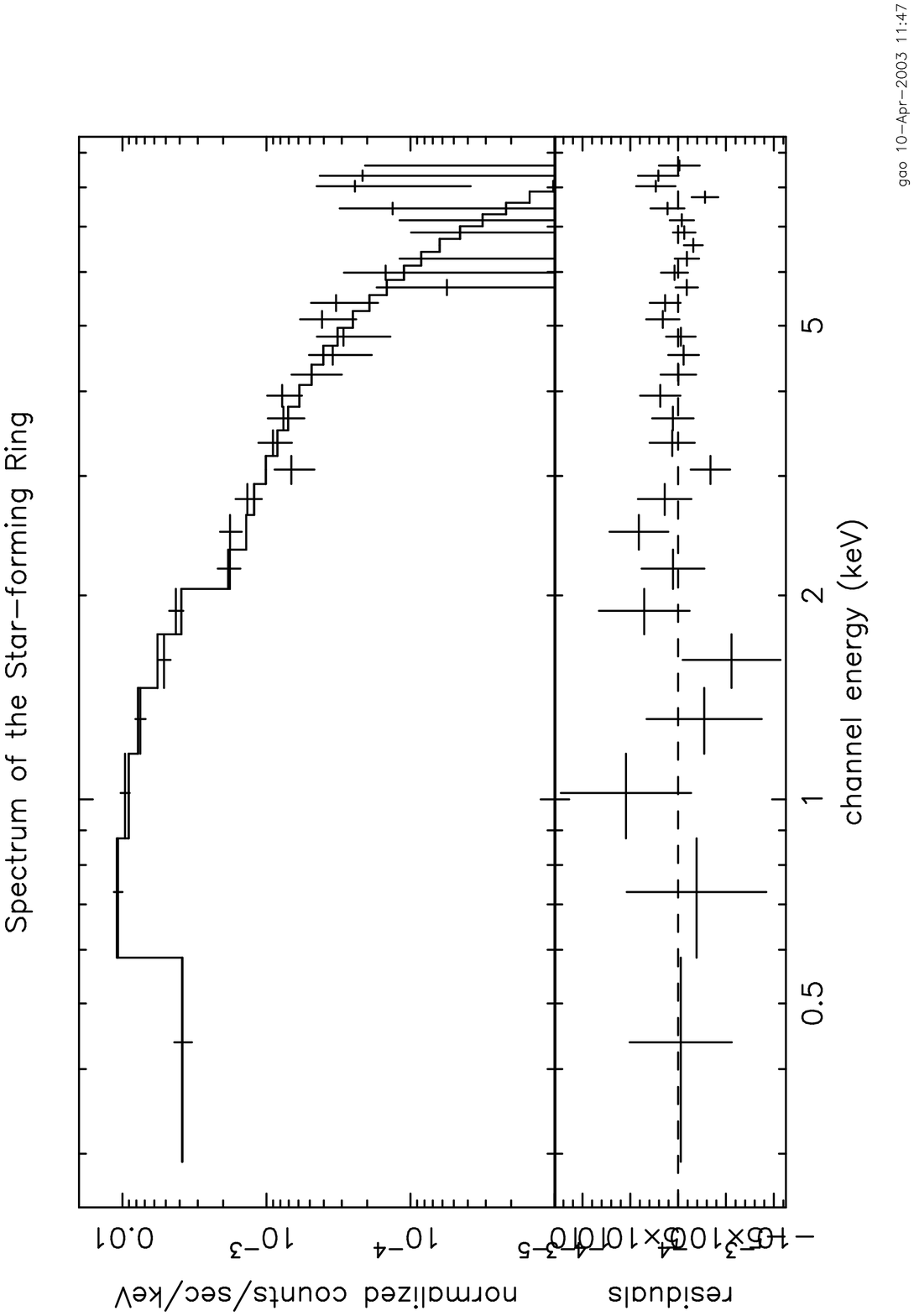}
\plotone{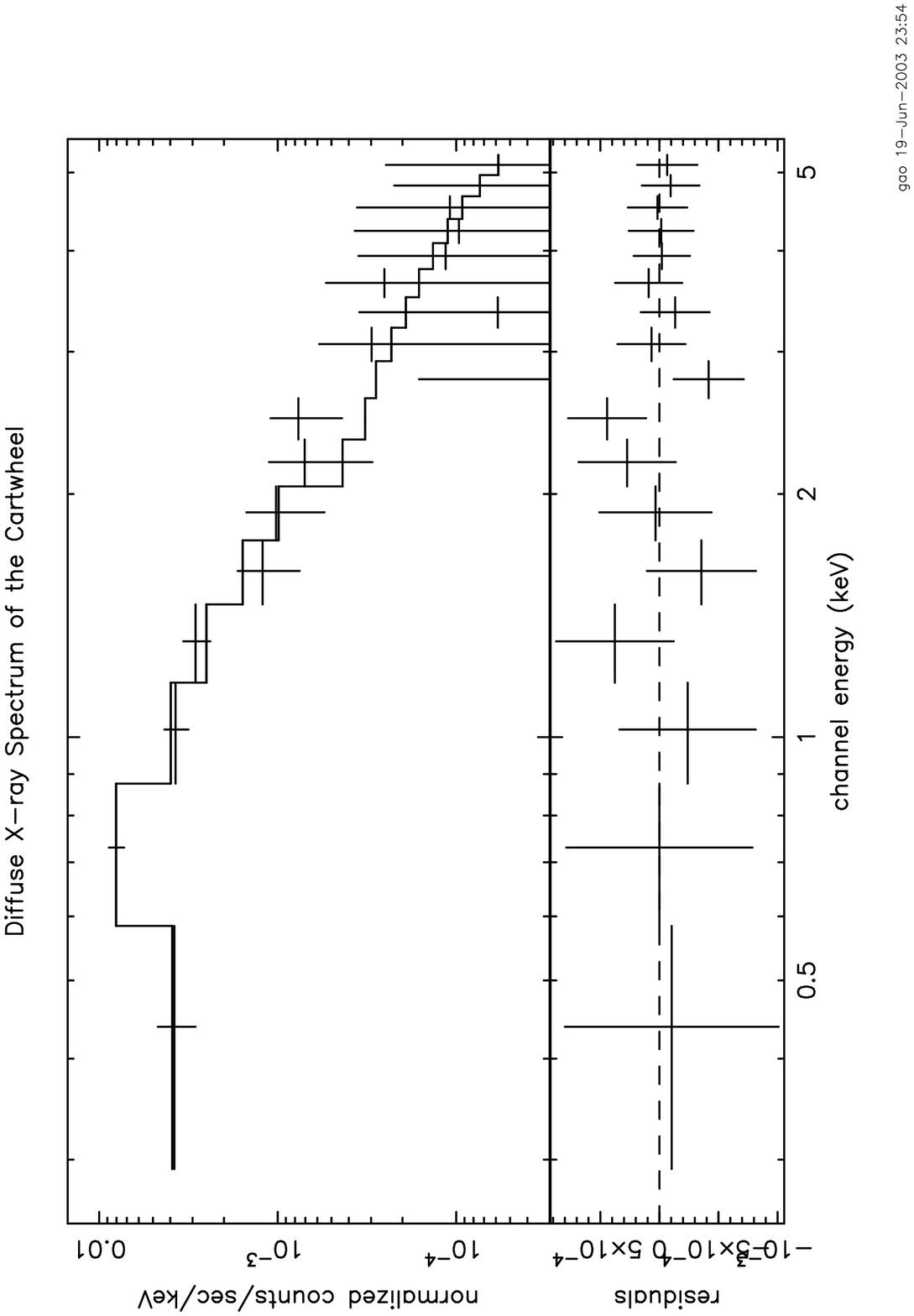}
\end{document}